\documentclass[prb,preprint,showpacs]{revtex4}

\usepackage{amsmath,amssymb}

\usepackage{graphicx}

\begin{document}

\widetext

\title{Can Quantum Lattice Fluctuations Destroy the Peierls Broken Symmetry Ground State?}

\author{William Barford$^{1}$ and Robert J. Bursill$^2$}

\affiliation{
$^1$Department of Physics and Astronomy, University of Sheffield, Sheffield, S3 7RH, United Kingdom\\
$^2$School of Physics, University of New South Wales, Sydney, New South Wales 2052, Australia
}

\begin{abstract}
The study of bond alternation in one-dimensional electronic systems has had a long history.  Theoretical work in the 1930s predicted the absence of bond alternation in the limit of infinitely long conjugated polymers; a result later contradicted by experimental investigations. When this issue
was re-examined in the 1950s it was shown in the adiabatic limit that bond alternation occurs for any value of electron-phonon coupling. The question of whether this conclusion remains valid for quantized nuclear degrees of freedom was first addressed in the 1980s. Since then a series of numerical calculations on models with gapped, dispersionless phonons have suggested that bond alternation is destroyed by quantum fluctuations below a critical value of  electron-phonon coupling. In this work we study a more realistic model with gapless, dispersive phonons. By solving this model with the DMRG method we show that bond alternation remains robust for any value of electron-phonon coupling.
\end{abstract}


\maketitle

The effect of quantizing the nuclear degrees of freedom on the bond
alternation was studied by Fradkin and Hirsch \cite{fradkin} on models of noninteracting
electrons. They used both renormalization group
arguments and Monte Carlo simulations in their investigations.  In a model with gapless, dispersive (or Debye) phonons they showed that for any value of the electron-phonon interaction the metallic state is unstable with respect to a
lower symmetry insulating phase that exhibits bond alternation.
This result confirms the adiabatic prediction.
The transition from the metallic state to the insulating state is called
the Peierls transition, with the ground state named the Peierls
state. They also predicted that for the same model with \emph{spinless
fermions} the Peierls state is destroyed by quantum lattice
fluctuations below a critical value of the electron-phonon interaction. Since a model of spinless fermions is equivalent to
the XY quantum antiferromagnet on open chains, this result
suggests that strong electronic interactions coupled to quantized
phonons might destroy the Peierls state.

More recent numerical calculations  on various models have also
indicated that the Peierls state may be destroyed by quantized
phonons below a critical value of the electron-phonon interaction.  Using a finite size scaling
analysis of the spin gap Caron and Moukouri \cite{caron} showed that at a critical value of the
electron-phonon interaction there is a Kosterlitz-Thouless
transition in the XY spin-Peierls model with  \textit{gapped,
non-dispersive} (or Einstein) phonons. Similar conclusions were
made for the Heisenberg spin-Peierls model, again with Einstein
phonons\cite{bursill99, uhrig}. The Holstein model with
Einstein phonons for both spinless\cite{bursill98} and
spinfull\cite{jeckelmann} fermions have also be shown to exhibit a Peierls
phase transition at a non-zero value of the electron-phonon
interaction.

Of more direct relevance to the present work is the investigation by Sengupta \emph{et al.}\cite{sengupta}
In this work the electronic degrees of freedom are modelled by a tight-binding model with onsite and nearest
neighbour Coulomb interactions (namely, the extended Hubbard model). The nuclear degrees of freedom are described by Einstein \emph{bonds} phonons. As in previous work with Einstein phonons, there is a critical value of the electron-phonon interaction below which lattice fluctuations destroy the Peierls state.

In this article we focus our attention on the role of quantized
\textit{gapless, dispersive} (or Debye) phonons on the Peierls
state.  The electronic degrees of freedom are modelled by the extended Hubbard
model, with the electronic and nuclear degrees of freedom being coupled
via linear distortions of the bond lengths. We study this model as
function of the electron-phonon interaction, the Coulomb
interaction and the phonon frequency. In all cases we consider a half filled band. In the limit of large
Coulomb interactions this model at half-filling maps onto the
quantum Heisenberg antiferromagnet, thus establishing a connection
with our previous work\cite{barford05}. Alternatively, in the
limit of vanishing phonon frequency (namely, $M \rightarrow
\infty$) the model maps onto the classical, adiabatic model, which
has been extensively studied (see, for example\cite{book} and
references therein).

The Hubbard-Peierls model with dispersive, gapless phonons is defined by,
\begin{eqnarray}\label{Eq:1}
    H =  H_1 + H_2
\end{eqnarray}
where
\begin{eqnarray}\label{Eq:2}
 H_1  = && -  \sum_{\ell=1, \sigma}^{N-1} \left( t + \alpha(u_{\ell} -
  u_{\ell + 1})\right)(c_{\ell \sigma}^{\dagger} c_{\ell+1
\sigma} + c_{\ell+1 \sigma}^{\dagger} c_{\ell \sigma})\nonumber
\\
&&  + U \sum_{\ell=1}^{N}
\left(N_{\ell\uparrow}- \frac{1}{2}\right) \left( N_{\ell\downarrow}-
\frac{1}{2}\right) + \frac{1}{2} \sum_{\ell} V (N_{\ell}-1)(N_{\ell+1}-1).
\end{eqnarray}
and
\begin{equation}\label{Eq:3}
H_2 = \sum_{\ell} \left(\frac{P_{\ell}^2}{2M} + \frac{K}{2}(u_{\ell} - u_{\ell + 1})^2\right).
\end{equation}
$u_{\ell}$ and $P_{\ell}$ are the conjugate displacement and
momentum operators for  the $\ell$th site.  $c^{\dagger}_{\ell
\sigma}$ creates an electron with spin $\sigma$ on site $\ell$.
$t$ is the electronic hybridization integral, $U$ and $V$ are the onsite and nearest neighbour Coulomb interactions, respectively, $\alpha$ is the electron-phonon coupling constant, $M$ is the
nuclear mass, and $K$ is the elastic spring constant.

In the adiabatic limit (defined by $M \rightarrow \infty$) and the
noninteracting limit (defined by $U = V = 0$) the bond
alternation, defined by
\begin{equation}\label{Eq:4}
\delta_0 = \frac{1}{N}\sum_{\ell} \frac{(u_{\ell +1}- u_{\ell})}{a}(-1)^{\ell}
\end{equation}
(where $a$ is the average bond length), satisfies
\begin{equation}\label{Eq:5}
 \delta_0 = 4\exp\left(-\left[1 +\frac{1}{2\lambda}\right]\right)
\end{equation}
for $\lambda \ll 1$. $\lambda$ is the usual definition of the
electron-phonon  coupling parameter, defined as $\lambda =
2\alpha^2/\pi Kt$. A non-zero value of the electron-phonon
interaction, $\lambda$, therefore implies a non-zero value for the
bond alternation. The purpose of this work is to address the role
of quantized phonons on the bond alternation, and in particular to
investigate whether there is a non-zero value of $\lambda$ for the
Peierls transition. We find that although quantized phonons reduce
the amplitude of the broken symmetry order parameter, in contrast
to the previous investigations using Einstein phonons, there is no
evidence that the Peierls state is destroyed at a finite value of
the electron-phonon interaction.

\section{Methods}

The lattice degrees of freedom are quantized for dispersive
phonons by  introducing the phonon creation and annihilation
operators,
\begin{equation}\label{Eq:6}
    b_{\ell}^{\dagger} = \sqrt{\frac{K}{\hbar\omega}}u_{\ell} -
    i\sqrt{\frac{\omega}{4 K \hbar}}P_{\ell}
\end{equation}
and
\begin{equation}\label{Eq:7}
   b_{\ell} = \sqrt{\frac{K}{\hbar\omega}}u_{\ell} +
    i\sqrt{\frac{\omega}{4 K \hbar}}P_{\ell},
\end{equation}
respectively, where $\omega = \sqrt{2K/M}$\cite{caron97,
barford02}.

Eq.\ (\ref{Eq:1}) is solved by the density matrix renormalization
group  (DMRG) method\cite{caron97,jeckelmann,zhang,barford02} for
chains of up to $136$ sites. The DMRG method is an efficient
truncation procedure for solving quantum lattice Hamiltonians,
especially in one-dimension\cite{white}. Solving electron-phonon
models, however, poses special problems as the number of phonons
is not conserved. Thus it is necessary to retain enough oscillator
levels per site ($M_p$) to ensure convergence of the key quantities. The number of
electron-phonon states per site ($4\times M_p$) is usually too
many to augment with the system block, thus it is necessary to
truncate this single site basis. There are therefore four
convergence parameters to tune for this model: the number of
oscillator levels per site, the number of optimized electron-phonon states
per site, the number of states per system block, and the overall
number of superblock states. 

We first establish convergence with respect to the number of optimized electron-phonon states per site. Table \ref{Ta:3} shows that both the ground state energy and the phonon order parameter (defined by Eq.\ (\ref{Eq:8})) have converged to better than $0.2\%$ for $14$ states per site. Next, we consider convergence with respect to the number of system and superblock states. Table \ref{Ta:4} shows excellent convergence when the number of system block states is $250$ and the eigenvalue product cutoff $\epsilon=10^{-14}$. Finally, we consider convergence with respect to the number of bare oscillator levels per site. Table  \ref{Ta:1} and Table  \ref{Ta:2} shows excellent convergence at $6$ levels per site for both the ground state energy and the phonon order parameter.

\begin{table}[tp]
\caption{The ground state energy, $E$, (in units of $t$) and the phonon order parameter, $Nm$, (defined by Eq.\ (\ref{Eq:8})) of the
Hubbard-Peierls model  as a function of the number of electron-phonon states per  site, $M$, for a
$40$-site chain with $6$ oscillator levels per site. $\omega =
t$, $U=2.5t$, $V=U/4$, and $g=0.1$.}
\begin{center}
\begin{tabular}{ccc}
\hline
\hline
$M$  & $E$ & $Nm$   \\
\hline
$6$ & $-58.716$ & $0.663$ \\
$10$ & $-59.969$ & $0.629$ \\
$14$ & $-60.220$ & $0.628$  \\
$18$ & $-60.275$ & $0.629$  \\
\hline
\hline
\end{tabular}
\normalsize
\end{center}
\label{Ta:3}
\end{table}

\begin{table}[tp]
\caption{The ground state energy, $E$, (in units of $t$) and the phonon order parameter, $Nm$, of the
Hubbard-Peierls model  as a function of the density matrix
eigenvalue product cutoff, $\epsilon$, the number of system block
states, $M$, and the superblock Hilbert space size, SBHSS, for a
$40$-site chain with $6$ oscillator levels per site. $\omega =
t$, $U=2.5t$, $V=U/4$, and $g=0.1$.}
\begin{center}
\begin{tabular}{ccccc}
\hline
\hline
$\epsilon$  & $M$ & SBHSS & $E$ & $Nm$ \\
\hline
$10^{-12}$ & $238$ & $141400$ & $-60.218$ & $0.641$ \\
$10^{-13}$ & $234$ & $232360$ & $-60.219$ & $0.633$ \\
$10^{-14}$ & $192$ & $285418$ & $-60.219$ & $0.637$ \\
$10^{-14}$ & $246$ & $359782$ & $-60.220$ & $0.629$ \\
$10^{-14}$ & $308$ & $415394$ & $-60.220$ & $0.629$ \\
$10^{-15}$ & $250$ & $506814$ & $-60.220$ & $0.628$ \\
$10^{-16}$ & $250$ & $652632$ & $-60.220$ & $0.629$ \\
\hline
\hline
\end{tabular}
\normalsize
\end{center}
\label{Ta:4}
\end{table}

\begin{table}[tp]
\caption{The ground state energy (in units of $t$) of the
Hubbard-Peierls model as  a function of the number of
sites, $N$. $\omega = t$, $U=2.5t$, $V=U/4$, and $g=0.1$.}
\begin{center}
\begin{tabular}{cccccccc}
\hline
\hline
  & \multicolumn{7}{c} \textrm{Number of oscillator levels per site} \\

$N$      &  $1$ & $2$  & $3$ & $4$ & $5$ & $6$ & $7$ \\
\hline
$16$  & $-23.072$ & $-23.506$ & $-23.635$ & $-23.668$ & $-23.684$  & $-23.685$ & $-23.685$ \\
$24$  & $-34.859$ & $-35.559$ & $-35.775$ & $-35.830$ & $-35.860$  & $-35.862$ & $-35.862$ \\
$40$  & $-58.440$ & $-59.671$ & $-60.058$ & $-60.162$ & $-60.215$ &  $-60.220$ & $-60.221$ \\
\hline
\hline
\end{tabular}
\normalsize
\end{center}
\label{Ta:1}
\end{table}

\begin{table}[tp]
\caption{The phonon order parameter, $Nm$, of the
Hubbard-Peierls model as  a function of the number of
sites, $N$. $\omega = t$, $U=2.5t$, $V=U/4$, and $g=0.1$.}
\begin{center}
\begin{tabular}{cccccccc}
\hline
\hline
  & \multicolumn{7}{c} \textrm{Number of oscillator levels per site} \\

$N$      &  $1$ & $2$  & $3$ & $4$ & $5$ & $6$ & $7$ \\
\hline
$16$  & $0$ & $0.321$ & $0.367$ & $0.367$ & $0.384$  & $0.385$ & $0.385$ \\
$24$  & $0$ & $0.398$ & $0.455$ & $0.455$ & $0.477$  & $0.479$ & $0.479$ \\
$40$  & $0$ & $0.518$ & $0.593$ & $0.595$ & $0.625$ &  $0.628$ & $0.629$ \\
\hline
\hline
\end{tabular}
\normalsize
\end{center}
\label{Ta:2}
\end{table}

Typically, we use $6$ oscillator
levels per site, with $14$ optimized single site states, $250$ block states,
and $500,000$  superblock states. One finite lattice sweep is
performed at the target chain size. In addition, \emph{in situ}
 optimization of the single site basis is performed for each site
 at the target chain\cite{barford02}.  In all cases we investigate linear chains with open boundary
conditions.
We maintain constant chains lengths by fixing the
position of the end sites.

The Hubbard-Peierls model in the classical limit (defined by Eq.\ (\ref{Eq:10})) is solved using the Hellman-Feynamn theorem which implies that,
\begin{equation}\label{Eq:12}
    (u_{\ell + 1} - u_{\ell}) = - \frac{\pi t \lambda}{\alpha}\left( \sum_{\sigma}\langle c_{\ell \sigma}^{\dagger} c_{\ell+1
\sigma} + c_{\ell+1 \sigma}^{\dagger} c_{\ell \sigma}\rangle - \Gamma \right),
\end{equation}
where $\langle \cdots \rangle$ means the expectation value in the ground state. The ground state bond alternation is found by iterating Eq.\ (\ref{Eq:12}) until there is convergence\cite{barford01, book}.

\section{Results and Discussion}

We first investigate the Hubbard-Peierls model for arbitrary
electron-phonon  interactions in the noninteracting limit ($U=V=0$). Following
Fradkin and Hirsch\cite{fradkin}, we use the dimensionless
electron-phonon interaction, $g = \sqrt{\alpha^2/Kt} \equiv
\sqrt{\lambda\pi/2}$.

The Peierls broken-symmetry ground state is characterized by a
non-zero value  of the staggered phonon order parameter, defined
by
\begin{equation}\label{Eq:8}
    m(N) = \frac{1}{N}\sum_{\ell}\langle B_{\ell+1}-B_{\ell}\rangle(-1)^{\ell},
\end{equation}
where $B_{\ell} = (b^{\dagger}_{\ell} + b_{\ell})/2$ is the
dimensionless  displacement of the $\ell$th site and $N$ is the
number of sites. Notice  from Eq.\ (\ref{Eq:6}) and Eq.\ (\ref{Eq:7})  that the bond alternation, $\delta_0$, is proportional to $\sqrt{\omega} m(N)$. To determine whether this order parameter
vanishes as a function of the electron-phonon interaction in the
asymptotic limit it is necessary to perform a finite size scaling
analysis. Fradkin and Hirsch\cite{fradkin} suggested that the
scaling relation
\begin{equation}\label{Eq:9}
    m(N) = \frac{1}{Ng}F(Nm({\infty}))
\end{equation}
should apply. Thus, curves of $Nm(N)$ versus $g$ are expected to
coincide  at the critical value of $g$ at which $m({\infty})$
vanishes.

\begin{figure}[tb]
\begin{center}
\includegraphics[scale=0.60]{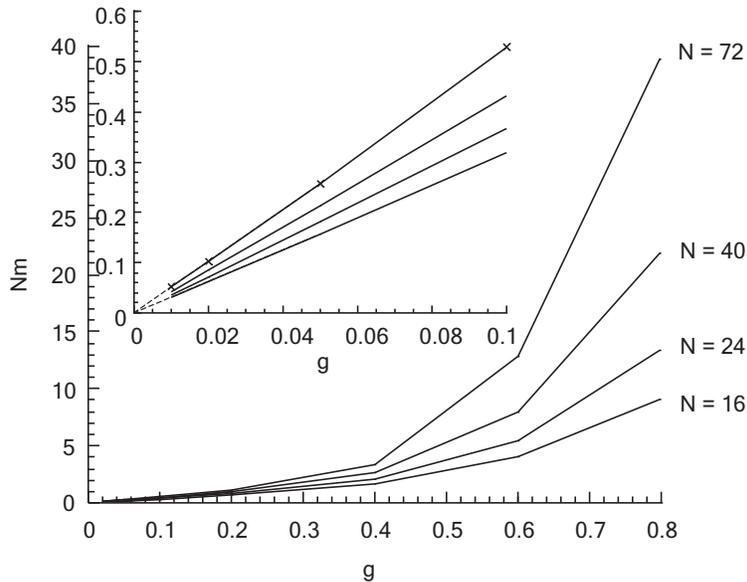}
\end{center}
\caption{The phonon order parameter versus the electron-phonon
interaction  for the noninteracting Hubbard-Peierls model. $\omega =
t$. The inset shows a linear extrapolation of $Nm$ versus $g$ to
the origin. (The crosses indicate the evaluated points.) } \label{Fi:1}
\end{figure}

Fig.\ (\ref{Fi:1})  shows $Nm$ versus $g$  for  values of $\omega = t$ and $U=V=0$. These results were obtained using the density matrix renormalization group method, as described in the Methods section. It is clear  that
for small values of $g$ $m$ is linearly proportional to $g$ for
all chain lengths. Furthermore, the curves extrapolate linearly to
the origin indicating a zero value for the critical $g$. This
confirms Fradkin and Hirsch's original prediction for the noninteracting limit of this model\cite{fradkin}, and accords with the adiabatic prediction. Notice
that since the distortion of the $\ell$th bond from its uniform
value, $(u_{\ell+1}-u_{\ell+1})$, is proportional to $\sqrt{\omega} g \langle
B_{\ell + 1} - B_{\ell} \rangle$, the  bond alternation is $\propto g^2$ or $\propto \lambda$ for $g \ll 1$.

We next consider the role of Coulomb interactions on the value of the bond alternation. It is known
that in the  adiabatic limit the bond alternation initially
increases as the Coulomb repulsion increases. This is because
electronic interactions suppress the quantum fluctuations between
the two degenerate bond alternation phases\cite{dixit, book}. The
bond alternation is maximized when the electronic kinetic energy
roughly equals its potential energy, namely when $U \sim 4t$. As the
Coulomb interaction increases further the bond alternation
decreases\cite{nakano, book}, eventually  scaling as $\sim t/(U-V)$.

\begin{figure}[tb]
\begin{center}
\includegraphics[scale=0.60]{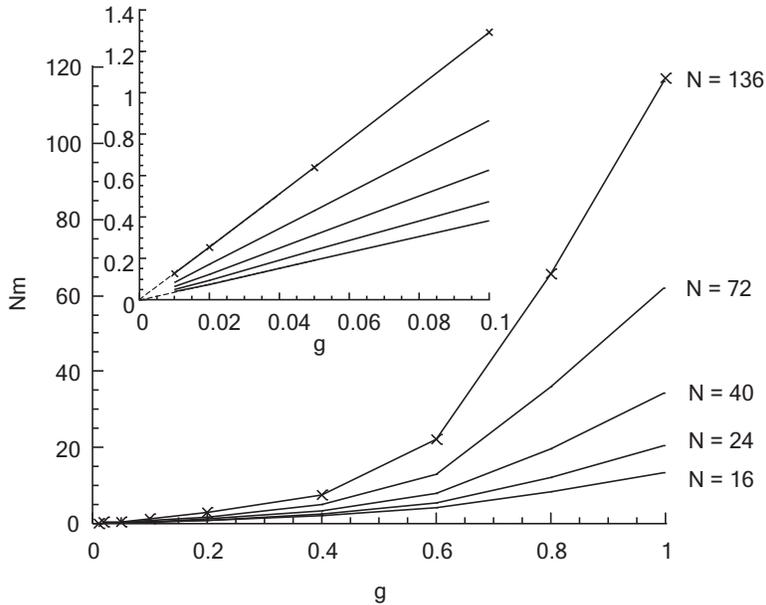}
\end{center}
\caption{The phonon order parameter versus the electron-phonon
interaction  for the interacting Hubbard-Peierls model. $\omega =
t$, $U=2.5t$, $V=U/4$. The inset shows a linear extrapolation of
$Nm$ versus $g$ to the origin. (The crosses indicate the evaluated
points.)} \label{Fi:2}
\end{figure}

Fig.\ (\ref{Fi:2})  shows $Nm$ versus $g$ for  values of $\omega = t$, $U=2.5t$, and $V=U/4$. These
Coulomb parameters were chosen to coincide with those used by
Sengupta \textit{et al.}\cite{sengupta}, who studied a similar
model to Eq.\ (\ref{Eq:1}), but with Einstein bond phonons. As already
stated, they reported that quantized phonons destroy the Peierls
ground state below a critical value of the electron-phonon interaction. In contrast to that work\cite{sengupta}, however, we see that for
our model with Debye phonons the predictions are qualitatively
similar to the noninteracting limit, namely $Nm$ decreases as a linear
function of $g$ as $g \rightarrow 0$. We note, however, that the
phonon order parameter for a given value of $g$ is larger for this value of $U$ than for
the noninteracting limit.

\begin{figure}[tb]
\begin{center}
\includegraphics[scale=0.60]{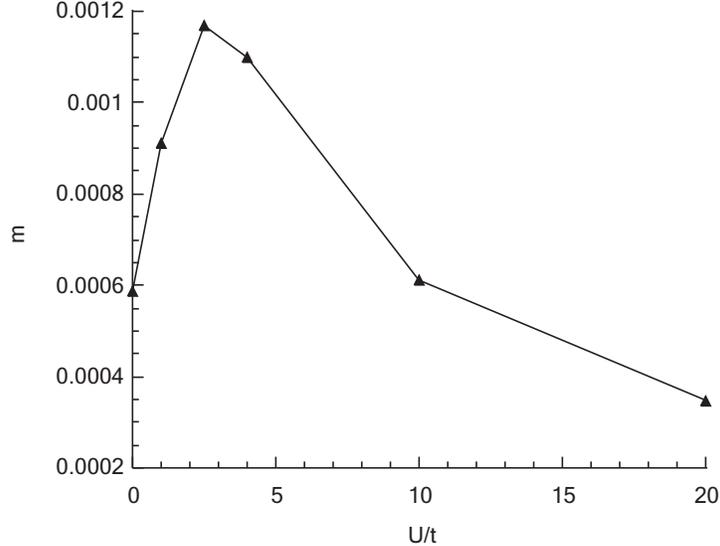}
\end{center}
\caption{The phonon order parameter for $72$-site chains versus the Coulomb interaction $U$ for the Hubbard-Peierls model. $\omega = t$, $V=U/4$, and $g=0.01$. } \label{Fi:5}
\end{figure}

The phonon order parameter, $m$, as a function of the Coulomb interaction $U$ is
illustrated in Fig.\ (\ref{Fi:5}) for $72$-site chains. $g = 0.01$, and thus the results lie in the linear regime where $m \propto g$. As for  the adiabatic limit, $m$
initially increases with $U$ before asymptotically decreasing as
$U \rightarrow \infty$. We might therefore expect that in the
strong coupling limit ($U \rightarrow \infty$) quantum
fluctuations have a better chance to destroy the Peierls state below a critical electron-phonon interaction. That this is not the
case was demonstrated by the present authors in an investigation of
the Heisenberg spin-Peierls model\cite{barford05}. This model is the
Hubbard-Peierls model in the strong coupling limit at
half-filling. The Heisenberg spin-Peierls model is parameterized
by $J \equiv 4t^2/(U-V)$, $\omega$ and $g$. Ref\cite{barford05}
showed that as for the Hubbard-Peierls model, $Nm$ also decreases as a linear function
of $g$ as $g \rightarrow 0$. We have therefore shown that for all
electronic interactions the Peierls state is robust to quantized
lattice fluctuations.

We now turn to discuss the phonon order parameter as a function of
the phonon frequency,  $\omega$. As $\omega \rightarrow 0$
(and more oscillator levels are occupied) the system approaches
the classical limit. To make a direct comparison to the classical
limit we also solve Eq.\ (\ref{Eq:1}) in this limit. The classical
Hamiltonian is defined by,
\begin{eqnarray}\label{Eq:10}
H = H_1 
+ \frac{K}{2}\sum_{\ell} (u_{\ell} - u_{\ell + 1})^2
+\Gamma\sum_{\ell}\left(u_{\ell} - u_{\ell + 1} \right)
\end{eqnarray}
(where $H_1$ is defined by Eq.\ (\ref{Eq:2})), and the bond variables are now
classical variables. The role  of the last term is to ensure 
constant chain lengths, where $\Gamma$ is self-consistently chosen
so that $\sum_{\ell}(u_{\ell} - u_{\ell + 1})=0$.
The ground state
bond alternation  of the classical model  is found iteratively  using
the Hellmann-Feynman procedure, as described in the Methods section.

\begin{figure}[tb]
\begin{center}
\includegraphics[scale=0.60]{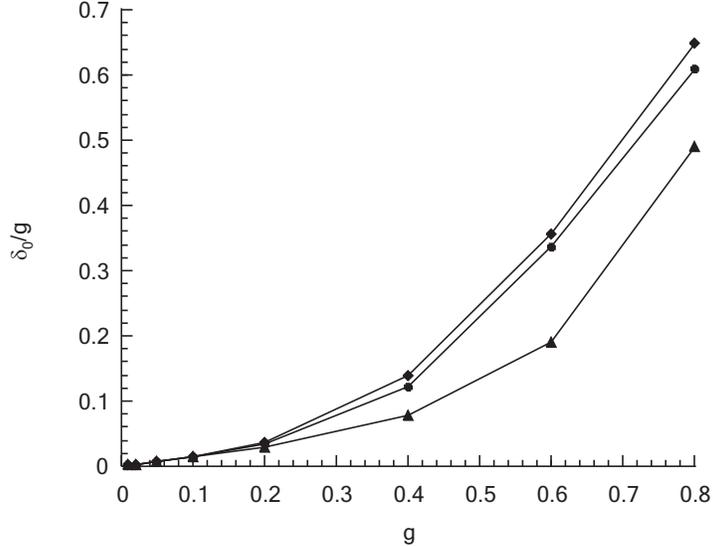}
\end{center}
\caption{The scaled bond alternation of the Hubbard-Peierls model for a $50$-site chain. 
The adiabatic limit (using Eq.\ (\ref{Eq:10}))  (diamonds); the quantum case (using Eq.\ (\ref{Eq:1})), $\omega = 0.1t$ (circles), and $\omega = t$ (triangles). $U = 2.5t$ and $V=U/4$.} \label{Fi:3}
\end{figure}

Fig.\ (\ref{Fi:3}) shows $\delta_0$ (scaled by $g$) of a $50$-site
chain versus $g$ for the  two quantum cases, $\omega = 0.1t$ and
$\omega = t$, as well as for the adiabatic limit. Evidently 
 $\delta_0/g$ tends linearly $\rightarrow 0$ as $g \rightarrow 0$.
Thus, $\delta_0 \propto g^2 \propto \lambda$. Moreover, as
predicted, the phonon order parameter decreases as a function of
$\omega$. As $g$ increases the deviation between the quantum and
classical predictions increases.

\begin{figure}[tb]
\begin{center}
\includegraphics[scale=0.60]{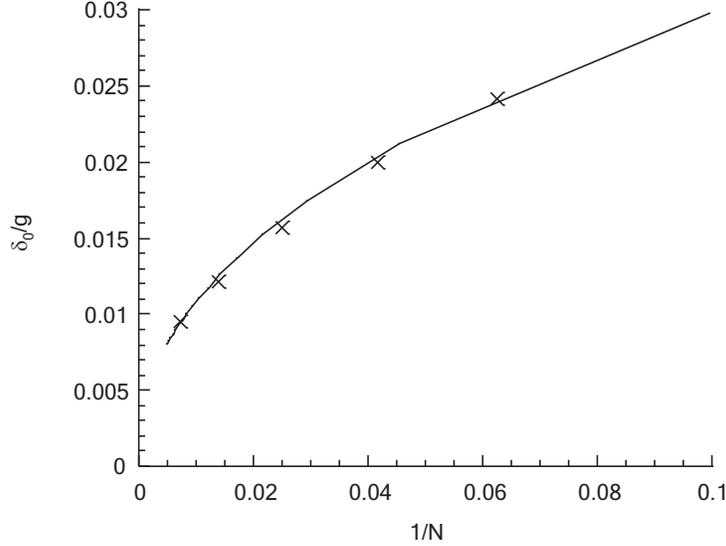}
\end{center}
\caption{The scaled bond alternation versus inverse chain length
in the adiabatic limit.  Crosses are the quantum result with
$\omega = t$. $g=0.1$, $U=2.5t$, and $V=U/4$.} \label{Fi:4}
\end{figure}

Finally, we discuss the nature of the Peierls transition at
$\lambda = 0$. In the  noninteracting limit the classical
transition is given by Eq.\ (\ref{Eq:5}). No analytical expression exists in
the interacting limit. However, Fig.\ (\ref{Fi:3}) shows that the
transition (at $g=0$) is qualitatively similar in the classical and quantum
cases, as $\delta_0$ is proportional to $g^2\equiv \lambda$. Moreover, the behaviour of the bond alternation as a
function of chain length is the same for small electron-phonon
interactions in the classical and quantum cases. Fig.\ (\ref{Fi:4}) shows the scaled bond alternation versus inverse chain length
in the adiabatic  and quantum cases. The close agreement of these predictions suggests that the transition at $g=0$ is the same in the quantum and classical cases in the limit of long chains.

\section{Conclusions}

In conclusion, we have used the DMRG method to investigate the
role of gapless, dispersive  quantized phonons on the  Peierls
transition in the Hubbard-Peierls model. 
We showed that the
staggered phonon order scales as $g$ (and the bond alternation
scales as $g^2$, and thus $\propto \lambda$) as $g \rightarrow 0$ for all values of the Coulomb interaction and phonon frequency considered here. In the strong coupling limit the Hubbard-Peierls model maps onto the Heisenberg spin-Peierls model, which also exhibits  a Peierls transition at $g=0$\cite{barford05}.
Thus, we conclude that the
Peierls transition occurs at $g=0$ for all values of the Coulomb interaction.
Moreover, we showed that the
quantum predictions for the bond order  follow the classical
prediction as a function of $1/N$ for small $g$. We therefore
conclude that the zero $g$ phase transition is of the mean-field
type.

\begin{acknowledgements}
W.\ B.\  thanks the Gordon Godfrey Bequest of the University of New South Wales, the Leverhulme Trust, and the Royal Society for financial support.
\end{acknowledgements}

The authors have no competing interests.

\end{document}